\documentclass{llncs}

\usepackage[utf8]{inputenc} 

\usepackage{booktabs} 
\usepackage{array} 
\usepackage{paralist} 
\usepackage{verbatim} 
\usepackage{subfig} 

\usepackage{amsmath}

\usepackage{marginnote}
\usepackage{xcolor}

\usepackage{url}

\usepackage{graphicx}
\usepackage{float}

\usepackage{mdframed}

\title{Electryo, In-person Voting with Transparent Voter Verifiability and Eligibility Verifiability}

\author{Peter B R\o nne\orcidID{0000-0002-2785-8301}, Peter Y A Ryan, Marie-Laure Zollinger} 
\institute{University of Luxembourg\\Esch-sur-Alzette, Luxembourg
\\\email{\{peter.roenne,peter.ryan,marie-laure.zollinger\}@uni.lu}}
\date{} 
         
\DeclareMathOperator{\ballotcode}{\textsf{bcode}}         
\DeclareMathOperator{\id}{\textsf{ID}}

\DeclareMathOperator{\sign}{\textsf{sign}}
\DeclareMathOperator{\pk}{\textsf{pk}}
\DeclareMathOperator{\sk}{\textsf{sk}}
\DeclareMathOperator{\vk}{\textsf{vk}}

\newcommand{\TT}{\mathit{TT}}
\newcommand{\TRA}{\mathit{TRA}}
\newcommand{\MN}{\mathit{M}}
\newcommand{\BB}{\mathit{BB}}
\newcommand{\EA}{\mathit{EA}}

\begin{document}

\maketitle

\begin{abstract}

Selene
is an e-voting protocol that allows voters to directly check their individual vote, in cleartext, in the final tally via a tracker system, while providing good coercion mitigation. This is in contrast to conventional, end-to-end verifiable schemes in which the voter verifies the presence of an encryption of her vote on the bulletin board. The Selene mechanism can be applied to many e-voting schemes, but here we present an application to the polling station context, resulting in a voter-verifiable electronic tally with a paper audit trail. The system uses a smartcard-based public key system to provide the individual verification and universal eligibility verifiability. The paper record contains an encrypted link to the voter’s identity, requiring stronger assumptions on ballot privacy than normal paper voting, but with the benefit of providing good auditability and dispute resolution as well as supporting (comparison) risk limiting audits.

\end{abstract}
\section{Introduction}

In this paper, we propose combining the highly transparent counted-as-intended verification mechanism of the e-voting scheme Selene \cite{selene} with paper ballot, in-person voting. The aim is to keep the vote casting experience close to paper ballot voting with optical scanning, while enabling the intuitive voter-verification of the Selene scheme. The resulting scheme provides improved dispute resolution and supports risk limiting audits.

For most end-to-end verifiable schemes the voter verifies the presence of an encryption of her vote in the input to the tally on the bulletin board. In contrast, Selene lets a voter check that her vote appears correctly, in the clear, in the final tally via a tracking number system. This provides a highly transparent and intuitive verification, but, if naively implemented, could lead to vote-selling and coercion. 

The main idea of Selene is to mitigate the coercion threats by notifying the voters of their tracking number only after the full list of tracking numbers and votes has been published.
Coerced voters can then simply choose a tracker pointing to the required vote and claim it as theirs.
The notification provides the voter high assurance that it is the correct, i.e. unique, tracker while being deniable in the event of coercion.

In a 
paper ballot election the voters enjoy ballot secrecy thanks to the isolation of the voting booth at the polling station - giving good resistance against coercion and vote-buying attempts. Normally, but with UK as a prominent counter-example, the ballots are also anonymous and unmarked, extending the ballot secrecy to the tally phase. However, the integrity of the election relies on trust assumptions for the talliers, and many real attacks and errors are known, as shown in \cite{errors}. 
In Germany, for example, the tally process is public \cite{germanelec} and at least gives the voters the possibility to oversee the tally ceremony, however considerable trust is still required in the chain of custody of ballots.

To improve on this situation we propose here introducing the Selene mechanism to allow voters to verify that their vote is counted-as-intended. This requires the introduction of a carefully protected link between the ballot and the voter.
The vote casting experience of the system is close to the optical scan paper ballot systems with the difference that the paper ballots will have a (QR-)code printed onto them which contains an encryption of the voter's identity.\footnote{This might be troublesome in some jurisdictions.} 
We assume that voters have smartcards to store and prove their ID.
Before getting into the details we recall the key elements of Selene.

\subsection{The Essence of Selene}

Selene revisits the old idea of enabling verification by posting the votes in the clear on the BB along with a private tracking numbers. The new twist is that voters are only notified of their tracker some time after the vote/tracker pairs have been publicly posted, giving a coerced voter the opportunity to choose an alternative tracker that will placate the coercer. Notification of the trackers is carefully designed to provide assurance that it is the correctly assigned tracker, i.e. unique to the voter, while being deniable. 
The key goals of Selene are:
\begin{itemize}
\item Ensure that each voter is assigned a unique tracker number.
\item Notify the voter of her tracker after the vote/tracker pairs have been published in a manner that provides high assurance and yet is deniable in the event of coercion.
\end{itemize}

This is achieved, in essence, by publishing a list of trackers, $n_i$, verifiably encrypting and shuffling these and assigning them to the voters under trapdoor commitments according to the secret permutation $\pi$ resulting from the shuffles. The commitment for the $i$th voter takes the form:

\begin{equation*}
\begin{aligned}
C_i:= \pk_i^{r_i} \cdot\, g^{n_{\pi(i)}}
\end{aligned}
\end{equation*}
Where $\pk_i$ is the voter's public trapdoor key. $C_i$ is a Pedersen commitment to the tracker but can also be thought of as the second term ($\beta$) of an exponential ElGamal encryption of the tracker under the $i$th voter's trapdoor public key $\pk_i$. The corresponding first term ($\alpha = g^{r_i}$) is not published, but is communicated to the voter over a private channel at notification time.  On receipt of the $\alpha$-term, the voter can combine this with the $\beta$-term and decrypt using her trapdoor key. 

If she is coerced, she can choose an alternative tracker that will satisfy the coercer and compute, using her trapdoor key, an alternative $\alpha$. 
Without the trapdoor, it is intractable to compute an $\alpha$ that will decrypt to a given tracker. This observation simultaneously underpins the assurance that the tracker is correct, and removes the need to authenticate the $\alpha$ as communicated to the voter.

\subsection{The Essence of Electryo}
The key innovation of Electryo is to introduce a protected link between the paper ballot and the voter ID, in such a way as to guarantee the integrity and the secrecy of the link. This link is used to associate the encrypted vote, scanned from the paper ballot, with the voter ID on the BB, thus enabling the Selene mechanism to kick in. An additional feature is that at the time of scanning the ballot, a fresh, random \emph{receipt code} is generated and printed for the voter to retain. This is required later to access the tracker number, providing an extra layer of privacy, as explained in detail later.

Now that voters are able to verify their vote in the clear, we can omit the usual checks required in cryptographic, end-to-end verifiable schemes: Benaloh challenges and correct posting to the $\BB$ of the encrypted vote. A corollary of this last observation is that the voter does not need to retain a copy of the encrypted vote, just the receipt code, which helps ensure receipt-freeness.

The voting system provides individual verifiability via the Selene check, allows universal verifiability of the setup phase and of the tally as well as eligibility via the digital signatures. The paper record provides a basis for dispute resolution, while risk-limiting audits will strengthen the link between the paper and digital record -- all of this while preserving a good measure of coercion-mitigation. 

The outline of the paper is as follows: Below 
we give a brief overview of related work. In section \ref{participants} we list the parties involved, as well as the primitives used. Section \ref{votingexp} will give details of the voting ceremony from the voter's point of view. In section \ref{protocol}, we will give further details of the scheme including the cryptography. Section \ref{analysis} gives a brief analysis of the scheme, describing some potential attacks and counter-measures.

\subsubsection*{Related Work}
Several in-person voting protocols mix paper ballots or a paper-audit trail with a public digital record of the votes:

Prêt-à-Voter \cite{pav} is a paper-based voting scheme with voter-verifiability, a version of which has been trialled in a state election \cite{DBLP:journals/corr/CulnaneRST14}. Contrary to the present scheme, these schemes does not provide transparent verification or directly support RLAs, see however \cite{DBLP:conf/esorics/LundinR08} for a version with a  human-readable paper-audit trail. 

Wombat \cite{DBLP:conf/ev/Ben-NunFLRRTW12} combines paper-ballot voting with cryptographic tabulation and end-to-end verifiability. A voting machine delivers a paper ballot containing a plaintext vote as well as the encrypted version as a QR-code. The voter can check the correctness of the plaintext vote before putting it in a ballot box. The encrypted version is scanned and posted to a $BB$ and the paper copy is kept by the voter as a receipt. 

Another polling station e-voting scheme is STAR-Vote \cite{starvote} which combines electronic voting machines (DREs) with a paper trail to achieve end-to-end verifiability and allow for efficient risk limiting audits (RLAs). The correctness of the encryption of the vote, can be tested by the voter by a sort of Benaloh challenge, where discarded ballots are decrypted in public. We will give a more detailed comparison in the long version of this paper (to appear). Note that it was not a design goal of STAR-vote to have eligibility verifiability.

All of these schemes do not provide the transparent verification of the plaintext vote in the final tally, as we do here.

We also note that there are other schemes based on trackers, specifically sElect \cite{select} and the boardroom scheme analysed in  \cite{CortierANR}, but they are  not paper-based and do not provide the level of coercion-resistance and receipt-freeness that we aim for here.
\section{Participants and Primitives} \label{participants}

The main participants of the protocols are:

\begin{itemize}
    \item The voters $V_i$. We assume that they are provided with electronic ID cards\footnote{See \cite{smartcardref} for an example of a smartcard implementing ElGamal encryption with Elliptic Curves.}, e.g. as part of a national electronic ID infrastructure like in Estonia \cite{estoniaweb}.\footnote{See \cite{2017-ccs-nemec} for a recently found flaw in that system, demonstrating the importance of a secure implementation of this system.} The card stores a secret signing key together with the ID which is associated with the corresponding public verification key $\vk_i$. We assume that the card can perform an encryption of the ID with the election key and sign input using the secret signing key. Further the voter has public and secret key pair, $\pk_i,\sk_i$ for the Selene mechanism, where the latter is stored in a Vote Supporting Device (VSD) e.g. a smartphone or a computer and perhaps also on the smartcard. 
    \item The Election Authority $\EA$ is managing the election and protocol setup.
    \item The Tally Tellers $\TT$ create the public election key $PK_T$ and threshold share the secret key. They also facilitate a Mixnet $\MN$ which is used to ensure privacy, and performs parallel verifiable re-encryption mixes see e.g. \cite{mixnet2}.
    \item A public web Bulletin Board $\BB$ is used for verifiable communication, and will be assumed to be append-only and have a consistent public view.
    \item The Tracker Retrieval Authority ($\TRA$) is responsible for relaying communication between the voters and $TT$. Specifically $\TRA$ will send the so-called $\alpha$ term to the voter, which can be turned into a tracker for her vote using the secret key $\sk_i$. 
    \item Registration Clerks and Talliers assisting at the polling station.
    \item Printers with Card Readers. These print a ballot code, $\ballotcode$, onto the paper ballots in the form of a QR-code, containing the re-encrypted ID and digital signature of the voter.
    \item Optical Scanners with a Receipt Printer. The Scanner reads out the voter's choice on the paper-ballot and the $\ballotcode$, and sends an encryption of the vote to $\BB$ together with a re-encryption of the ballot code and an encrypted receipt code. It delivers a \emph{ballot proof} to the voter, that contains the receipt code in plain text together with a digital signature for accountability. 
\end{itemize}

Some primitives used are
\begin{itemize}
    \item Encryption. We assume an IND-CPA secure homomorphic encryption scheme allowing re-encryption and verifiable mixing. To be explicit we choose
    ElGamal encryption which was used in Selene, and the homomorphic properties are needed for the Selene mechanism. We denote encryption with the key $PK$ $\{\cdot\}_{PK}$ and re-encryption is denoted $\{\cdot\}'_{PK}$. For some parts the homomorphic properties are not necessary and we use a RCCA secure scheme instead, i.e. the only malleability of the ciphertext is the ability to re-encrypt which is necessary for privacy and mixnets. To be explicit we can use the OAEP 3-round transformation \cite{DBLP:conf/asiacrypt/PhanP04,DBLP:conf/fc/PereiraR17} of ElGamal. 
    A single ciphertext then basically consists of two ElGamal ciphertexts and is RCCA secure under the Gap Diffie-Hellman assumption. We denote this encryption $\{\cdot\}_{\textrm{OAEP},PK}$. The parallel mixing is easily adapted to this encryption scheme since it basically consists of two ElGamal ciphertexts.

    \item Zero-Knowledge Proofs. We use zero-knowledge proofs, and proofs of knowledge, as well as signatures to ensure universal verifiability.  
    For non-malleability the strong form \cite{DBLP:conf/asiacrypt/BernhardPW12} of the Fiat-Shamir transform \cite{DBLP:conf/crypto/FiatS86} is used for obtaining non-interactive proofs, and we further include election identifiers in the hash to avoid malleability across elections.

    \item Plaintext equivalence tests (PETs). A PET \cite{DBLP:conf/asiacrypt/JakobssonJ00} produces a public verifiable test whether two ciphertexts are an encryption of a same plaintext message, without revealing the plaintexts to anybody. The test requires a threshold set of the Tellers $\TT$. 

    \item QR-codes. A QR-code is a matrix barcode containing information for reliable and easy scanning. 
    The encryption schemes used here can be based on elliptic curves requiring in the order of 512 bit strings. A ciphertext could then e.g. be stored in a QR-code version 6 (up to 1088 bits) or a version 10 for two OAEP ciphertexts.
\end{itemize}
\section{Voting Experience} \label{votingexp}

\begin{figure}[ht]
\centering \includegraphics[width=10cm]{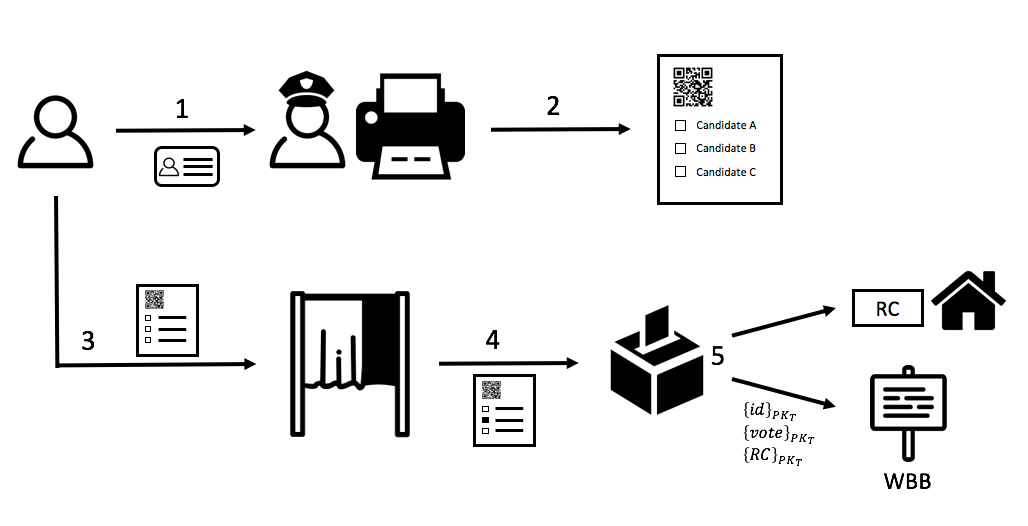}
\caption{Description of the voting phase\label{fig:voting}.\\
(1) The voter enters the polling station and goes to a registration clerk with her ID card to be identified. (2) Her ID card is read and the printer delivers the ballot with the encrypted ID contained in a QR-Code. (3) The voter goes to a booth to fill her ballot. (4) She puts her ballot into a ballot box containing the scanner, (5) that sends the encrypted vote to the bulletin board and prints the voter a take home receipt code.}
\end{figure}

In this section we describe the protocol from the voter's perspective. The vote-casting ceremony is close to a paper-ballot election with optical scanning of the ballots. The entire voting experience is described in figure \ref{fig:voting} and more cryptographic details will be given in next section.

\subsection{Registration}

We assume that all voters are in possession of ID smart-cards, e.g. as part of a national 
electronic ID infrastructure. The ID card can create signatures and will be used to authenticate voters. The registration of the voters could probably happen automatically if based on a national PKI, alternatively by a company etc.
Besides the ID-card, the Selene mechanism assumes that each voter $V_i$ holds a secret key, $\sk_i$. 
This may require a registration step by the voters, the details of which we omit, but note that these keys could potentially be used for multiple elections.
The tracker authority $\TRA$ also needs to know where to contact the voter for the tracker retrieval phase, e.g. an email address. Such confirmed contact data is normal to have in an electronic ID infrastructure. However, for improved usability, we assume that the voter is using an app (e.g. authenticated via the sim card as in Estonia \cite{estoniaweb}) that accesses $\sk_i$ e.g. via the smartcard or, if properly authenticated, the program could be the creator of the Selene keys. 

\subsection{Voting phase}

On the voting day, the voter presents her ID card to a poll worker to confirm ID and eligibility, as in standard elections the voters showing up can be recorded in a paper log. The printer is equipped with a smart card reader and interacts with the card to retrieve an encryption, by the smart-card, of the ID and signature. It prints an unfilled ballot with a QR-code which encodes a re-encryption of the voter ID and digital signature, confirming the voter was present. 

Then the voter enters the booth to fill the ballot, and finally she heads to a ballot box with a scanner/printer. The latter delivers a receipt code $RC_i$ on paper, without releasing it, before scanning the ballot.
The scanner re-encrypts the ballot code, encrypts the vote and releases the receipt code to the voter. The data is stored and sent to $\BB$ after voting ends, and the paper ballot is retained in the ballot box.

\subsection{Tracker retrieval}

After the tally phase, cast votes and corresponding tracking numbers will appear on the $BB$. After a pause, allowing coerced voters to access this information, the voters will receive their $\alpha$-term (see introduction) via their support device at randomised times, as in Selene. The device will calculate the voter's unique tracker using the received $\alpha$-term, the public $\beta$-term and the trapdoor key $\sk$.

However, in contrast to Selene the $\alpha$ term will only be sent to the voter if she at some point after election enters a correct receipt code $RC_i$ in her device. This adds a layer of privacy as explained later in section \ref{sec:ballotprivacy}.

\subsection{Voting in case of coercion}

Coerced voters can take steps to mitigate the coercion.

After the tally board is created with votes and corresponding trackers, the voter can choose a tracker pointing to a candidate of the coercer's choice. Further, the voter can calculate a fake $\alpha$-term using $\sk$ that opens to this tracker. The voter can now show the coercer this tracker and $\alpha$-term, if required. 

Further, for improved coercion-resistance, the coerced voter can also contact $\TRA$ with authentication and request to not receive the real $\alpha$-term, but only the fake. Now, even in the case where a coercer or vote buyer controls the interface to receive the $\alpha$-term, he does not receive any convincing evidence of the cast vote. As mentioned above the essential assumption here is that the voter has access to $\sk$, e.g. via multiple copies or the storage on the ID card. 

In a longer version of this paper (to appear), we will present an alternative version of Selene where the coerced voter even before or during voting can contact $\TRA$ and request a faked $\alpha$-term.

\subsection{Comments on usability}

The voting experience is close to a standard optical-scan scheme.
As with an optical-scan or STAR-Vote \cite{starvote}, the scanner and ballot box can be combined, so that the ballot will be read before being fed in automatically in the ballot box.
The only aspect that might be a bit troubling for some voters is the printing of the QR code on the ballot form. This does not affect usability, it is automatic as far as the voter is concerned, but might be worrying from a privacy perspective.

We avoid the verification steps such as Benaloh challenges \cite{DBLP:conf/uss/Benaloh06} of the encryptions. Instead, we have the extra Selene verification phase with the receipt code and tracker check, which we believe is more understandable for voters.

For disabled persons, multi-lingual communities or generally complicated ballots, voting machines could also be used to fill out the ballots. Here the QR code created by the printer is scanned by the voting machine to produce the filled out ballot, which is kept as a paper record. A scanning step is not necessary in this case. 
\section{Protocol Description} \label{protocol}

We now describe the protocol in more technical detail including cryptography.

\subsection{Pre-Election Setup}

Let us recall Selene's set-up phase that we will also follow here \cite{selene}.

The Tally Tellers set up a secure group and create the threshold election key $PK_T$ for ElGamal encryption (or another homomorphic encryption scheme).\\
We assume that all voters have PKs in the chosen group. Let $\pk_i = g^{x_i}$ be the public key of voter $V_i$, and $x_i=\sk_i$ their secret key. The Election Authority publishes on $BB$ the set of tracking numbers $n_i$. These could just be $1,\ldots,n$ with $n$ the number of eligible voters.  
Using a verifiable re-encryption mix each voter is associated a unique secret encrypted tracker on $\BB$: $(\id_i,\{g^{n_j}\}_{PK_T})$, where $j:=\pi(i)$, and $\pi$ is the secret permutation resulting from the mixes.
As described in detail in \cite{selene}, the Tally Tellers $TT_1,\ldots,TT_t$ produce a trapdoor commitment $C_i = \pk^{r_i}_i \cdot g^{n_{j}}$ where $r_i = \sum_{k=1}^{t} r_{i,k}$, along with an $\alpha$-term $\alpha_i=g^{r_{i}}$ that will be kept secret under encryption.
Only $TT_k$ knows $g^{r_{i,k}}$.\footnote{A difference to \cite{selene} is that we do not introduce separate Tracker Tellers, but instead let the Tally Tellers handle this, and we introduce a single separate Tracker Retrieval Authority $\TRA$.}\\
Before vote casting $\BB$ displays
$$(\id_i, \vk_i, \pk_i,  \{g^{n_{j}}\}^{\prime}_{PK_T}, C_i)$$

Here $\vk_i$ is the verification key for voter $V_i$, and the corresponding secret key is stored along with $\id_i$ on the voter's smartcard. The smartcard can produce signatures that can be verified via $\vk_i$ and we assume the signature scheme to be existentially unforgeable. Further, the smartcard can produce encryptions that can be used in the mixnet construction and decrypted by $\TT$. We here use ElGamal encryption $\{\cdot\}_{PK_T}$, the OAEP version thereof discussed above, and e.g. Schnorr signatures, but other choices are possible, and since the smartcards are used across elections it might be preferable to use a separate key for this part.

\subsection{Voting}

Voter $V_i$ goes to the polling station and is identified and registered by a clerk. If her identity is confirmed and if she has not voted yet, the clerk proceeds to the printing. 
The ID card 
is read and delivers an encryption of the voter's ID to the printer. The latter re-encrypts it (to avoid privacy attacks from a colluding ID card and scanner) and delivers a QR-code representing the ballot code $\ballotcode=(\{\id_i\}_{\textrm{OAEP},PK_T}, \{\sign_i\}_{\textrm{OAEP},PK_T})$.\footnote{Cryptographically it would suffice to leave out the encryption of $\id_i$, since it can be determined from testing different $\vk$'s.} The signature is of the ID and the election ID, but can also include e.g. the date and the printer ID. The clerk should be screened from seeing the printed ballot, but can check that the correct ID card is read in the card reader.

After retrieving her ballot, the voter enters a booth and fills out the ballot with her vote $vote_i$ by hand.

The voter now proceeds to a ballot box that contains a scanner. The scanner first prints a receipt code, that is not yet detachable from the ballot box. This ensures that the receipt code does not depend on the vote and thus cannot be used as a subliminal channel. The receipt code is a random short pin, e.g. five digits, with check digits. The voter then puts her ballot in the box, the scanner reads it and releases the receipt code. It processes the data and re-encrypts the ballot code elements, encrypts the vote and receipt code, and publishes on $BB$ (after election, if offline):
$$\{\id_i\}'_{\textrm{OAEP},PK_T}, \{\sign_i\}'_{\textrm{OAEP},PK_T}, \{\ballotcode\}_{PK_T}, \{\textrm{vote}_i\}_{PK_T}, \{RC_i\}_{PK_T},\Pi_i$$
Here $\Pi_i$ is a zero-knowledge proof of plaintext knowledge for the vote and receipt code and correct message space, for less malleability we suggest to include an AND-proof, proving that the two first encryptions are re-encryptions of the $\ballotcode$ in the third ciphertext. We include the election identifier in the hash of the Fiat-Shamir transform. The proofs will prevent vote copy attacks also across elections. The reason to re-encrypt the ciphertexts in the ballot code is to prevent coercion attacks via taking a picture of the filled-in ballot as a proof of the cast vote. In this case, a coerced voter can fill out a ballot as required by the coercer, photograph it, and go back to the officials for a new ballot and hand the (photographed) one, which is destroyed. They now cast their intended vote using the new ballot form. The re-encryption means that the paper ballot won't be linkable to the public electronic record, which is also important in the RLAs.
Finally, $\{\ballotcode\}_{PK_T}$ is an encryption of the ballotcode which is here written in shorthand, but includes several ElGamal ciphertexts. If needed, these can be decrypted and allow crosschecking with the corresponding paper record.

\subsection{Mix and decryption}

These published tuples are now sent through a parallel mixnet (e.g. Verificatum \cite{mixnet2}) on the BB after checking the proofs $\Pi_i$. After decryption of the first term we get back the $\id$ and signature, i.e. we get mixed tuples of

$$\id_i,\sign_i,\{\ballotcode\}'_{PK_T},\{\textrm{vote}_i\}'_{PK_T}, \{RC_i\}'_{PK_T}$$

Now the signature can be checked for eligibility verification and with the previous data on the $BB$ we construct (suppressing re-encryption for clarity)

$$\id_i, \{g^{n_{\pi(i)}}\}_{PK_T}, C_i, \{\textrm{vote}_i\}_{PK_T} $$

As in Selene, the second and last term, the encrypted tracker and vote, are put through a verifiable parallel mix, after which the Tellers perform a verifiable decryption, to obtain the final tally board containing tracker/vote pairs:
$$(n_{\pi(i)}, \textrm{vote}_i)$$

\subsection{Tracker notification}

\paragraph{Receipt verification} 
Before she can check her vote, the voter must enter the receipt code $RC_i$ on her device (after log in). The app will encrypt the receipt code and a $TT$ will do a PET between this encryption and the one displayed on the bulletin board.
This verification is also done to ensure that the paper ballot and the corresponding electronic record are related to the same voter, i.e. to prevent an attack from the printer putting the wrong ID on the ballot, see section \ref{analysis} for details.

\paragraph{Tracker retrieval} The public commitment $C_i$ and the corresponding $\alpha$-term can be combined to form an encryption of the tracker under the voter's public key: 

$$(\alpha_i,C_i) = (g^{r_i}, \pk^{r_i}_i \cdot g^{n_{\pi(i)}})$$

If the voter has entered the correct receipt code, the $\alpha_i$ term will be sent to the voter, and she can then compute the decryption using her secret key and retrieve
$g^{n_{\pi(i)}}$, and hence her unique tracker $n_{\pi(i)}$.
The Tracker Retrieval Authority will get the $\alpha_{i}$ shares from each Tally Teller (authenticated for accountability), multiply these together to obtain $\alpha_{i}$ and send this unauthenticated to the voter.

As described in Selene \cite{selene} it is computationally hard, without knowing $\sk_i$, to calculate an alternative $\alpha$-term that opens to a valid tracker. Thus the $\alpha$-terms can be transmitted unauthenticated to the voter. On the other hand the voter can efficiently calculate such a fake $\alpha$-term for any tracker (see \cite{selene}), and thus shows this to a coercer in case of coercion.

\subsection{Risk Limiting Audits}

A comparison Risk Limiting Audit (RLA) \cite{rla} is a method to confirm (or refute) the outcome of an election to any required confidence, by random sampling of the paper ballots. The digital and paper records of the vote are compared. Typically, for reasonably large margins, a small sample will suffice to achieve a good level of confidence, e.g. 95\%. This technique requires a link between the digital and paper copies for every ballot. 

The RLA testing, 
can be used to monitor the behaviour of the scanners. The audit should be performed in both directions, i.e. first start from tuples on $\BB$, decrypt the $\ballotcode$ and find the corresponding paper ballot and check the consistency. In the other direction we can also start from a paper ballot, and the corresponding encrypted ballot can be found via PET tests, or more efficiently via an obfuscation of a part of the ballot code by lifting to a secret power homomorphically and then decrypting. 
\section{Preliminary Security Analysis} \label{analysis}

In this section, we give an overview of the security properties and their corresponding assumptions, and a brief, informal analysis of potential attacks. A more rigorous, formal analysis will be the topic of future work.

As is standard with E2E V schemes, we assume a trustworthy $\BB$ giving a consistent view for universal verifiability. This might be implemented as, for example \cite{vvotewbb}, or perhaps using some form of Distributed Ledger Technology.

\subsection{Verifiability}

The current scheme has strong emphasis on verifiability and integrity, and like STAR-vote it has triple assurance by producing three lines of evidence for the result, firstly via tally board for the electronic ballots obtained via secure mixnets, secondly via the Selene mechanism that lets voters check their cast vote in the clear in the tally. Thirdly, the paper record enables comparison RLA checks between the paper and electronic records of the encrypted votes. Furthermore, the plaintext votes in the paper ballots can be hand counted if required.
We also have universal eligibility verifiability via the ID-system, and the ID-marked paper ballots provide a solid base for dispute resolution.

The system remains secure even under large-scale collusions: even if the $\BB$ is malicious and can simulate views different views to voters, an honest RLA on paper ballots can detect a manipulated result. On the other hand, if all parties are corrupted including auditors, but the $\BB$ remains trustworthy and the Selene secret trapdoor keys are not compromised (e.g. stored on malware-free devices), then voters who check their vote can be sure it is counted correctly (assuming a computational assumption see \cite{selene}).

\paragraph{Universal Verifiability} Universal verifiability allows voters and even third parties to verify the correct tallying of the votes on the $\BB$. Anyone, including an interested voter, can check the proofs of the verifiable mixing and decryption of the encrypted votes and trackers. Further, the cryptographic operations on the $\BB$ during the setup phase of the Selene mechanism can to be verified by any observer. This will guarantee that each voter is assigned, and later notified, a unique tracker
(assuming that the voter's Selene (trapdoor) keys are kept secret). As is true in general for e-voting schemes, the checks done by an individual voter are not enough to assure her that the overall outcome is correct. This is because she needs to be confident that all the other votes have also been correctly recorded and counted. This is assured if sufficient numbers of other voters have also checked their votes, or other mechanisms guarantee the correct handling of the votes. Here we have the possibility of auditors checking that the paper record is consistent with the digital record of encrypted votes on the $\BB$. As an additional feature, as in STAR-vote, the interested party can also follow the evidence output by the RLAs that puts a risk-limit on the final result.

\paragraph{Eligibility Verifiability}

Eligibility is firstly checked by a registration clerk at the polling station, maintaining a log of attending voters. However additionally, the signature, produced by the smartcard and included in the ballot code, can be publicly verified on the bulletin board -- even by third parties. Assuming secrecy of the signing keys, ideally protected with a pin on the smartcard, this proves the presence of the voter due to the existential unforgeability of the signature scheme, and prevents ballot stuffing. If the ID cards are used often, one concern might be that an adversary before the election maliciously obtains a signature of the data to be signed at the election. This can be prevented if the election ID contains some public random data first obtained briefly before the election.

Note that the tracker retrieval implicitly checks that the voter ID appeared on $\BB$ associated with some ballot, and then the check of the tracker on the tally board confirms correct vote. However the presence of their ID can also be checked independently by voters not using the Selene mechanism, or people abstaining. And anybody can check that only valid IDs and corresponding signatures appear giving the universal eligibility verifiability assuming an honest ID-card system.

\paragraph{Individual Verifiability} Individual verifiability ensures that voters can check that their vote is recorded as intended. Typically, this is done by allowing a voter to challenge the encryption of the vote and then check the cast ciphertext on $\BB$ and then relying on the universal verifiability for the correct inclusion of the vote in the tally. The Selene mechanism used here, on the other hand, gives a more transparent direct verification directly of the cast vote in plaintext in the final tally. We hope that this will help incentivise more voters to do the check.

Attacks in the spirit of the trash attack \cite{trash} might appear troublesome. If the printer could see an $\id$ associated to a voter, that is assumed not to perform any checks, it could then choose the ballot code (e.g. the last digits) to give instructions to a colluding scanner that it can maliciously change the electronic output vote. That is one reason, to let the ID card encrypt the ID, such that the printer won't be able to know the voter's identity. This attack thus requires a collusion between the ID card, printer and scanner and to identify a voter not doing the check. Further, the adversary also needs to collude with a tallier to miscount a vote in order to not get a discrepancy between the paper and online record. Further the attack could be caught by the RLAs. We see that the adversary needs a large collusion and has a risk of detection, whereas in normal paper voting the adversary only needs to control the tallier.

As with all E2E V schemes, the verification performed by an individual voter provides assurance that her vote is correctly handled but does not of itself provide assurance that the outcome of the election is correct. For that the voter needs to be confident that all other votes have been correctly handled. Usually this depends on a reasonable number of voters being sufficiently diligent in performing the appropriate checks, in particular checking that their encrypted vote appears correctly on the $\BB$. This is perhaps a questionable assumption, but here, as with \cite{DBLP:conf/esorics/LundinR08}, we have an independent (paper) record of the cast encrypted votes which allows independent auditors to supplement the voter checks, lessening the dependence on voter diligence. This, along with the RLAs on the plaintext votes, should provide ample assurance to all that the outcome is correct.

Another attack vector is malware infection of the VSD for the Selene check, however it still requires a large collusion to actually change a vote.

\subsection{Dispute resolution} 
One of the issues with the original (internet) Selene scheme was how to resolve a complaint when a voter claims that the vote they find on the $\BB$ is not the vote that they cast. If this occurs it is important for a judge to be able to determine where the complaint is genuine, arising from a corruption or malfunction of the system, or just a belligerent or forgetful voter. 
In Electryo, the election authorities can, in camera, request the corresponding ballot code on $\BB$ to be decrypted. The corresponding paper ballot can be identified using the ballot code allowing the vote, the ID and the corresponding signature can be checked.

\subsection{Ballot privacy}\label{sec:ballotprivacy}

Whereas we have high security guarantees for verifiability, the price for this has to be paid in terms of stronger assumptions for privacy. Ballot privacy essentially means that the voting system should not reveal a non-negligible amount of information about an honest voter's vote other than the outcome of the election. For a detailed overview of game-based definitions see \cite{DBLP:conf/sp/BernhardCGPW15}, and also \cite{smythintro} for a recent work. 

Clearly we need to assume that at least one Mix Teller is honest for ballot privacy, and that we do not have a threshold set of corrupted Tally Tellers. Also, if the ID card, the printer and the scanner collude, they can trivially map votes to IDs. However, if one of these are honest, then the adversary cannot link the IDs to the plaintext votes.

Further, each voter's supporting device is trusted for the privacy of the corresponding vote, but only if the voter actually uses the mechanism and enters the receipt code. This gives an extra layer of security for voters with high privacy concerns and who do not wish to trust the device and perform the check. It also safeguards voters with less technical knowledge, for whom the adversary has managed to setup and control the Selene keys and device. Note also that even if the secret Selene trapdoor key is leaked, the adversary still needs access to the $\alpha$-term to break privacy.

The check of the additional receipt code via a PET is also used to guard against an attack where the adversary uses the Selene mechanism itself to check a vote, i.e. we must make sure that the Selene check is directed to the correct voter. Consider the case where the printer and smartcard are colluding, but the scanner is honest. To find the vote of the attacked voter, the printer uses the ID and signature key for a colluding voter and prints this on the ballot. Via the Selene mechanism this colluding voter can learn the vote of the attacked voter. This would be detectable via a Selene check done by the attacked voter, however the check of the receipt code means that this attack is detected with high probability and before the privacy leak happens, since the malicious voter does not know the correct $RC_i$.

The scanner itself can also carry out attacks on privacy by delivering a wrong output. Consider the case where the scanner sees an interesting vote, and wants to know the corresponding ID. It could then try to put the ciphertext of the ID in two different output tuples, and after the mixing and decryption of the IDs and signatures, it searches for two identical IDs. Likewise it could also change the ciphertext of the signature to produce an invalid signature, and then look for this on $\BB$. Finally if the PETs of the receipt-codes are not done privately it could also provoke an error here, by encrypting a false $RC_i$. However, all of these attacks are immediately detectable with overwhelming probability and will deter a risk-adverse adversary from doing it. The RCCA security of the ciphertexts of the ID and signature means that it cannot do more advanced attacks than replacing or copying ciphertexts. The marking of ciphertexts used in \cite{DBLP:conf/fc/PereiraR17} 
could also be introduced to make sure that the ciphertexts come from the ballot printer. Another worry could be that the scanner uses ciphertexts of votes from earlier elections, where they are publicly related to the IDs. However the ciphertexts in the output needs to come with proofs of plaintext knowledge, so the IND-CPA security prevents attacks of this kind.

Let us finally turn to the paper record. A problem here arises if a registration clerk manages to see the ballot code at registration, and is able to memorize part of it and to communicate this to a colluding tallier to break privacy. This is a reason to use QR-codes that are hard to memorise for humans, and we use printers with a shielding screen such that the clerk cannot see the printed ballot. In standard paper ballot voting, it is also not hard to make almost undetectable marks on the ballot. Alternatively we speculate that high resolution pin-hole cameras could be used by adversaries to photograph the ballots (backlit) before and after voting. The privacy could then be broken using that paper have a unique fiber structure (see e.g. \cite{DBLP:conf/wifs/Wong015}). We can never do better than the ordinary paper ballot scheme in this respect.

\subsection{Receipt-Freeness and Coercion-Resistance}

The receipt-freeness of a voting system says that the voting system should not give the voter any evidence to prove to a third party how she voted. The Selene e-voting protocol was carefully designed not to give such a receipt.
However, it was only designed for use  in low level coercion settings.

Firstly, the talliers of the paper ballots are trusted for receipt-freeness since the voter could make an almost invisible mark the ballot, a mark which the tallier will then look for. Likewise the scanner also needs to be trusted for receipt-freeness, since a malware could also be triggered with such a mark, something similar applies to the DRE machine of STAR-Vote, where a certain input pattern could alert the malware. However, the scanner could also do more direct attacks, it could e.g. store the mapping between receipt values and votes. This could be repaired by having a separate scanner of the QR code, and let this print the receipt code, and publish the encryptions on $\BB$.

During the verification phase the $\TRA$ is trusted since it directly knows the true $\alpha$-terms. The Tally Tellers however only need to be trusted to not be colluding in a threshold way which would break privacy. 

We have also made sure that the ballot code only appears (re)encrypted on $\BB$, and that the voters does not have a receipt showing the $\ballotcode$.

The RLAs can also endanger receipt-freeness if not performed carefully, e.g. the ID should not be revealed in plain, but only the correct links via plaintext-equivalence tests. Also Italian attacks can be countered, see \cite{DBLP:conf/uss/BenalohJLLS11LONG}.

On the other hand, side channels attacks might occur. A coercer might be sufficiently convinced by a simple ballot selfie, and today little is done to actually guard against these. 
Alternatively it could also be printed after vote-casting on the filled-in ballot that have been folded to hide the voter's choice.

Finally note that a symbolic model of Selene has been proven receipt-free \cite{DBLP:conf/voteid/BruniDS17} for the case where we use auxiliary trackers per voter, 
see the longer version of this paper (to appear), where it is also shown how to request fake trackers before the end of the voting phase. 
Standard Selene suffers from a mild lack of coercion-resistance from the case where the coerced voter chooses a fake tracker which is the adversary's.

\section{Conclusion}

In this paper we propose a polling station voting protocol that combines a paper ballot record with a digital record to provide individual verifiability via the transparent Selene mechanism, universal verifiability and universal eligibility verifiability. Further the paper ballots are marked with the encrypted ID of the voters and provide a strong basis for dispute resolution and risk-limiting audits. However, as privacy and integrity are dual requirements, the strong integrity guarantees, especially for eligibility, comes at the cost of stronger assumptions for ballot privacy.

The hope is that the integrity guarantees and the transparent individual checks and good dispute resolution properties can be used to create trusted election results e.g. in cases where the integrity of earlier elections have been questioned.

For future work, the scheme certainly needs a full security model and corresponding analysis and proofs. Here it would also be natural to use automated verification in a simplified symbolic model, to make sure that attacks have not been overlooked. Cryptographically it would be interesting if the scheme could be improved such that we can move to a stronger version of the covert adversary model for privacy. Right now the adversary can pull off a privacy attack, but it will be detected. We could introduce a second authority which checks the decrypted IDs and signatures on $\BB$ before publishing these, such that the adversary would not be able to gain any knowledge from the attempted privacy attack. However, for integrity complicated zero-knowledge proofs are necessary when decryptions are not published, and efficient versions needs to be found. 

Another important aspect for future research is the usability and user experience of the scheme, which should also include the officials. 

Finally some variants of the scheme would be interesting to explore. Firstly, a postal version of the scheme. Postal voting is an increasingly popular voting form, but with integrity problems that should be addressed. Also, it would be interesting to explore a version of the scheme with everlasting privacy or participation privacy, but maybe milder guarantees for the universal part of the eligibility verifiability. Further, combining Pr{\^{e}}t {\`{a}} Voter with the Selene check might have some advantages in privacy compared to the present scheme. Finally, the trusted platform problem is an important issue for the individual verification, and it would be interesting to explore the use of an extra device for the Selene check, e.g. utilizing the ID-card and a card reader.

\subsection*{Acknowledgements}
We would like to thank the Luxembourg National Research Fund (FNR) for funding, in particular PBR was supported by the FNR INTER-Sequoia project which is joint with the ANR project SEQUOIA ANR-14-CE28-0030-01, and MLZ was supported by the INTER-SeVoTe project.

\bibliographystyle{splncs04}
\bibliography{selene_ps}

\appendix

\end{document}